\documentclass[12pt,aps,onecolumn,floats,superscriptaddress,twoside,pre,Letter]{revtex4}
\usepackage{graphicx}
\usepackage{amsmath}
\usepackage{amssymb}
\usepackage{mathtext}

\begin{document}

\title{Assessment of the dependence of $(\delta{S}/\delta{V})$ on the heat influx for a well-stirred two-phase system with interfacial boiling}

\author{Denis S.\ Goldobin}
\affiliation{Institute of Continuous Media Mechanics, UB RAS,
             Perm 614013, Russia}
\affiliation{Department of Mathematics, University of Leicester,
             Leicester LE1 7RH, UK}
\affiliation{Perm State National Research University,
             Perm 614990, Russia}
\author{Anastasiya V.\ Pimenova}
\affiliation{Institute of Continuous Media Mechanics, UB RAS,
             Perm 614013, Russia}

\begin{abstract}
For a well-stirred multiphase fluid systems the mean interface area per unit volume, $(\delta{S}/\delta{V})$, is a significant characteristic of the system state. In particular, it is important for the dynamics of systems of immiscible liquids experiencing interfacial boiling. We estimate the value of parameter $(\delta{S}/\delta{V})$ as a function of heat influx $\dot{Q}_V$ to the system.
\end{abstract}

\maketitle

\section{Introduction}

For a well-stirred multiphase fluid systems the mean interface area per unit volume, $(\delta{S}/\delta{V})$, is an important characteristic of the state. It becomes even more significant for the systems where this interface is active chemically or in some other way. The systems of immiscible liquids experiencing interfacial boiling~\cite{Krell-1982,Geankoplis-2003,Simpson-etal-1974,Celata-1995,Roesle-Kulacki-2012-1,Roesle-Kulacki-2012-2,Sideman-Isenberg-1967,Kendoush-2004,Filipczak-etal-2011,Gordon-etal-1961,Prakash-Pinder-1967-1,Prakash-Pinder-1967-2,Pimenova-Goldobin-JETP-2014,Pimenova-Goldobin-2014-2} are an example of the systems where parameter $(\delta{S}/\delta{V})$ becomes especially important. The problem of calculation of $(\delta{S}/\delta{V})$ cannot be addressed rigourously and any direct numerical simulation, being extremely challenging and CPU-time consuming, will provide results pertaining to a quite specific system set-ups. Some generale assessments on $(\delta{S}/\delta{V})$ can be highly beneficial. In this paper we perform these assessments for the process of direct contact boiling in a system of two immiscible liquids.

At the direct contact interface, a vapour layer grows and produces bubbles which breakaway of the interface and rise. The presence of vapour bubbles change the fluid buoyancy and perform a ``stirring'' of the system. This stirring enforces increase of the contact area $S$, while surface tension and gravitational segregation of two liquids counteract the increase of the contact area.

For a two-liquid system experiencing direct contact boiling, the quantity of our interest depends on parameters of liquids and characteristics of the evaporation process, which are controlled by the mean overheating and the bubble production rate~\cite{Filipczak-etal-2011}. In this work the volumes of both components are assumed to be commensurable, no phase can be considered as a medium hosting dilute inclusions of the other phase. The characteristic width of the neighborhood of the vapour layer, beyond which the neighborhood of another vapour layer lies, is
\begin{equation}
\nonumber
H_1+H_2\sim\left(\frac{\delta{S}}{\delta{V}}\right)^{-1}.
\end{equation}
The relationship between $H_1$ and $H_2$ is
$$
\frac{H_1}{H_2}=\frac{\phi_1}{\phi_2}=\frac{\phi_1}{1-\phi_1}\,,
$$
where $\phi_j$ is the volumetric fraction of the $j$-th liquid in
the system. It will be convenient to use
\begin{equation}
H_j\sim\phi_j\left(\frac{\delta{S}}{\delta{V}}\right)^{-1}.
\nonumber
\end{equation}

The process of boiling of a mixture above the bulk boiling temperature of the more volatile liquid is well-addressed in the literature~\cite{Simpson-etal-1974,Celata-1995,Roesle-Kulacki-2012-1,Roesle-Kulacki-2012-2,Sideman-Isenberg-1967,Kendoush-2004,Filipczak-etal-2011}. Hydrodynamic aspects of the process of boiling below the bulk boiling temperature~\cite{Pimenova-Goldobin-JETP-2014,Pimenova-Goldobin-2014-2} has to be essentially similar at the macroscopic level; rising vapour bubbles drive the stirring of system, working against the gravitational stratification into two layers with a flat horizontal interface, the surface tension forces tending to minimize the interface area, and viscous dissipation of the flow kinetic energy. Specifically, the behaviour of parameter $(\delta{S}/\delta{V})$ depending on macroscopic characteristics of processes in the system should be the same as for systems with superheating of the more volatile component. In what follows, we perform an analytical assessment of the dependence of $(\delta{S}/\delta{V})$ on the evaporation rate (or heat influx) for a well-stirred system.

\section{Energy flux balance in a well-stirred system}
Let us attempt to derive the rough relationships between the macroscopic parameter $(\delta{S}/\delta{V})$ of the system state and the heat influx rate per unit volume $\dot{Q}_V=\delta{Q}/(\delta{V}\delta{t})$ for a statistically stationary process of interfacial boiling.

The flow and consequent stirring in the system are enforced by the buoyancy of the vapour bubbles, while other mechanisms counteract the stirring of the system. These other mechanisms are gravitational stratification of two liquids, surface tension tending to minimise the interface area and viscous dissipation of the flow energy. Since the latent heat of phase transitions and heat of temperature inhomogeneities are enormously large compared to the realistic values of the kinetic energy of microscopic motion and gravitational potential energy~\footnote{Indeed, the energy of thermal motion of atoms corresponds to characteristic atom velocities $10^2-10^3\,\mathrm{m/s}$, while nothing comparable can be imagined for macroscopic flow velocities in realistic situations. The latent heat of water evaporation is even significantly bigger than the kinetic energy of thermal motion of its atoms at $T=300\,\mathrm{K}$.}, the latter can be neglected in consideration of the heat balance. Hence, all the heat inflow into the system can be considered to be spent for the vapour generation;
 $\dot{Q}_VV\longrightarrow(\Lambda_1n_1^{(0)}+\Lambda_2n_2^{(0)})\dot{V}_v$,
where $V$ is the system volume, $\dot{V}_v$ is the volume of the vapour produced in the system per unit time, $\Lambda_j$ is the enthalpy of vaporization per one molecule of liquid $j$, and $n_j^{(0)}$ is the saturated vapour pressure of liquid $j$. Thus,
\begin{align}
\dot{V}_v=\frac{\dot{Q}_V\,V}
 {\Lambda_1n_1^{(0)}+\Lambda_2n_2^{(0)}}\,.
\label{eq-apb-01}
\end{align}

The potential energy of buoyancy of rising vapour bubbles $\rho_lV_vgh/2$ (where $h$ is the linear size of the system, $h\sim V^{1/3}$, $\rho_l$ is the average density of liquids, the vapour density is zero compared to the liquid density) is converted into the kinetic energy of liquid flow, the potential energy of a stirred state of the two-liquid system, the surface tension energy and dissipated by viscosity forces. In a statistically stationary state, the mechanical kinetic and potential energies do not change averagely over time and all the energy influx is to be dissipated by viscosity;
$$
\rho_lV_vgh/2\longrightarrow\dot{W}_{l,k}\tau\,,
$$
where $\dot{W}_{l,k}$ is the rate of viscous dissipation of energy, $\tau$ is the time of generation of the vapour volume $V_v$, $V_v=\dot{V}_v\tau$. Hence,
\begin{align}
\rho_l\dot{V}_vg\frac{h}{2}\sim\dot{W}_{l,k}\,.
\label{eq-apb-02}
\end{align}

Let us estimate the viscous dissipation of the kinetic energy of flow $W_{l,k}$;
\begin{align}
\dot{W}_{l,k}&=\int\limits_V\vec{v}\cdot\vec{f}_\mathrm{vis}\mathrm{d}V
 \sim\int\limits_V\vec{v}\cdot\left(-\eta_l\frac{\vec{v}}{H^2}\right)\mathrm{d}V
\nonumber\\
 \sim& -\frac{\eta_l}{\rho_l}
 \frac{2}{\left(\frac{H_1+H_2}{2}\right)^2}
 \int\limits_V\frac{\rho_lv^2}{2}\mathrm{d}V
 \sim-8\nu_l\left(\frac{\delta{S}}{\delta{V}}\right)^2W_{l,k}\,.
\label{eq-apb-03}
\end{align}
Here $\vec{v}$ is the liquid velocity, $\vec{f}_\mathrm{vis}$ is the viscous force per unit volume, $H$ is the spatial scale of flow inhomogeneity, which is the half-distance between the sheets of the folded interface between liquid components, $\eta_l$ and $\nu_l$ are the characteristic dynamic and kinematic viscosities of liquids, respectively.

Further, we have to establish the relationship between the flow kinetic energy and the mechanical potential energy in the system. Rising vapour bubbles pump the mechanical energy into the system, while its stochastic dynamics is governed by interplay of its flow momentum and the forces of the gravity and the surface tension on the interface. In thermodynamic equilibrium, the total energy is strictly equally distributed between potential and kinetic energies related to quadratic terms in Hamiltonian (this statement is frequently simplified to a less accurate statement, that energy is equally distributed between kinetic and potential energies associated with each degree of freedom). Being not exactly in the case where one can rigorously speak of thermalization of the stochastic Hamiltonian system dynamics, we still may assess the kinetic energy of flow to be of the same order of magnitude as the mechanical potential energy of the system. Thus,
\begin{align}
W_{l,k}\sim W_{l,pg}+W_{l,p\sigma}\,,
\label{eq-apb-04}
\end{align}
where $W_{l,pg}$ and $W_{l,p\sigma}$ are the gravitational potential energy and the surface tension energy, respectively. We set the zero levels of these potential energies at the stratified state of the system with a flat horizontal interface.

The gravitational potential energy of the well-stirred state with uniform distribution of two phases over hight is
$$
W_{l,pg}\sim\Delta\rho_lVg\frac{h}{2}\,,
$$
where $\Delta\rho_l$ is the component density difference.
The surface tension energy is
$$
W_{l,p\sigma}\sim(\sigma_1+\sigma_2)V\left(\frac{\delta{S}}{\delta{V}}\right)\,,
$$
where we neglected the interface area of the stratified state compared to the area $V(\delta{S}/\delta{V})$ in the well-stirred state. Due to the presence of the vapour layer between liquids the effective surface tension coefficient of the interface is $(\sigma_1+\sigma_2)$ but not $\sigma_{12}$ as it would be in the absence of the vapour layer.

\section{The value of $(\delta{S}/\delta{V})$ yielding balance of energy fluxes}

Collecting Eqs.~(\ref{eq-apb-01})--(\ref{eq-apb-04}), one finds
\begin{align}
\rho_l\frac{\dot{Q}_V\,V}{\Lambda_1n_{1\ast}^{(0)}+\Lambda_2n_{2\ast}^{(0)}}g\frac{h}{2}
\approx
 8\nu_l\left(\frac{\delta{S}}{\delta{V}}\right)^2
 \left[\Delta\rho_lVg\frac{h}{2}+(\sigma_1+\sigma_2)V\left(\frac{\delta{S}}{\delta{V}}\right)\right]\,.
\nonumber
\end{align}
This equation can be simplified to
\begin{align}
\dot{Q}_V\approx
 B\left(\frac{\delta{S}}{\delta{V}}\right)^2
 \left[1+\frac{2}{k_{12}^2h}\left(\frac{\delta{S}}{\delta{V}}\right)\right]\,,
\label{eq-apb-05}
\end{align}
where $B=8\nu_l(\Lambda_1n_1^{(0)}+\Lambda_2n_2^{(0)})\Delta\rho_l/\rho_l$ and $k_{12}=\sqrt{(\rho_2-\rho_1)g/(\sigma_1+\sigma_2)}$. Noteworthy, the relative importance of the first and second terms in the brackets in Eq.~(\ref{eq-apb-05}) depends on the system size $h$.

For the n-heptane--water system, $B\approx1.5\,\mathrm{J/(m\cdot s)}$ and $l_{k_{12}}\equiv1/k_{12}\approx0.5\,\mathrm{cm}$~\cite{Pimenova-Goldobin-2014-2}. For a well-stirred system the distance between sheets of the folded interface $(\delta{S}/\delta{V})^{-1}\ll h$. The average compound of these two values can be either small or large compared to $l_{k_{12}}$;
\\
(1)~$h\cdot(\delta{S}/\delta{V})^{-1}\ll l_{k_{12}}^2$ corresponds to the case of the surface tension dominated system,
\\
(2)~$h\cdot(\delta{S}/\delta{V})^{-1}\gg l_{k_{12}}^2$ corresponds to the case of a gravity-driven system.

Cubic equation~(\ref{eq-apb-05}) possesses only one positive solution which is real-valued for any value of $\dot{Q}_V/B$;
\begin{align}
\left(\frac{\delta{S}}{\delta{V}}\right)
 =\left(\frac{\delta{S}}{\delta{V}}\right)_g\cdot
 G\left(\frac{\big(\dot{Q}_V/B\big)^{1/2}}{k_{12}^2V^{1/3}}\right),
\label{eq-apb-06}
\end{align}
where $(\delta{S}/\delta{V})_g=(\dot{Q}_V/B)^{1/2}$ is the value of parameter $(\delta{S}/\delta{V})$ for a gravity-driven system and function $G(s)=(6s)^{-1}(R+R^{-1}-1)$, $R=(\sqrt{27}s+\sqrt{27s^2-1})^{2/3}$; $G(0)=1$ and $G(s\gg1)=(2s)^{-1/3}$.

\section{Conclusion}
Basing on the energy flux balance in the system and assumption of the system stochastization, we have assessed the value of $(\delta{S}/\delta{V})$. Expression~(\ref{eq-apb-06}) allows estimating the value of parameter $(\delta{S}/\delta{V})$ as a function of heat influx $\dot{Q}_V$ to the system.

The work has been financially supported by the Russian Scientific Foundation (grant no.\ 14-21-00090).

\end{document}